\documentclass{rrparticle}
\usepackage{geometry}

\usepackage[utf8]{inputenc}
\usepackage{latexsym,amsmath,multicol,booktabs,calligra}
\usepackage{graphicx,listings,stackengine}
\usepackage{float}
\usepackage{amsmath}
\usepackage{amsthm}
\usepackage{eucal}
\usepackage{amssymb}
\usepackage{mathrsfs}

\title{ THE DOG-AND-RABBIT CHASE PROBLEM \\  \scriptsize {Updated (typos eliminated+appendix) version of paper R. A. Gutoiu, T. O. Cheche, Rom. Rep. Phys., 74(1), 903, 2022 }}

\author[1]{R. A. Guțoiu}
\author[1*]{T. O. Cheche}
\affil[1] {University of Bucharest, Faculty of Physics, P.O.Box MG-11, RO-077125, Bucharest – Măgurele, Romania}

\affil[1*]{Corresponding author Email: tiberius.cheche@unibuc.ro}
\date{July 2021}

\begin{document}
\maketitle

\begin{abstract}
The dog-and-rabbit chase problem is a classical problem that illustrates the concepts of elementary kinematics and, therefore, can be used in introductory mechanics teaching. By dealing with the relative motion, the problem naturally requires solving elementary differential equations and stimulates students to learn mathematics. The efficiency of the approach we adopt is emphasized by extending the class of problems that can be solved by applying it.

\end{abstract}

\section{Introduction}
Learning and teaching physics in the university requires a gradual introduction, as a level of complexity, of specific topics and subjects. On another hand, mathematics is indispensable in this knowledge process, and teaching it should also be gradual. And last but not least, the programming and use of scientific software programs become an important component in the study of science. All these three parts of the curriculum are considered in physics education targeting their application in education and research.
\par
In a science study program, the differential equation theory is one of the primary topics the university students learning physics are dealing with. The relativity of motion is a topic that crosses all fields of physics, while C++, Fortran programming languages or Mathematica software (only to give a few examples) are valuable research tools.
\par
A large number of works dealing with differential equations, and which treat the relativity of motion, and use programming may be listed. Mentioning a few such leading works and some related works of one of the authors might be useful to readers.
Thus, the strain field in heterogeneous elastic structures \cite{1}  is modeled by solving analytically or numerically systems of differential equations by using programs developed by the authors in Fortan and Mathematica \cite{2,3,4}. For the absorption and photoluminescence spectra of GaAs/AlAs and InAs/AlAs in semiconductor quantum dots, the quantum optical rate equations are simulated \cite{5,6,7}. To calculate the intrinsic spin Hall conductivity the Hamiltonian of a two-dimensional electronic gas is introduced in the differential time-independent Schröddinger equation \cite{8, 9,10,11}, while to simulate the electron transfer reaction rates and relaxation in dissipative systems the Nakajima–Zwanzig integral equations are solved \cite{12,13}. In the study of dynamics of a pulsejet engine in vertical motion in a uniform gravitational field without \cite{14} and with drag \cite{15}, convective velocity in the atmospheric layers\cite{16}, or the probabilistic coin toss modeling \cite{17,18} and deterministic complex motion of the gravitational pendulum  \cite{19}, the models are based on differential equations which are solved with codes written in Python and/or Mathematica.
\par
In the present pedagogical work we consider two classical problems of kinematics which addresses the relative motion. Both consider two inertial systems of reference and a particle that moves respecting different scenarios in each of the problems. The innovative character consists in observing that a careful choice of the system of references and coordinates allows a generalization of the solving method, which in turn leads to the definition of a class of problems. The strategy we adopt is to solve the two related problems in parallel and find their common core.
\par
The paper is structured as follows. In section II the two problems are introduced. The characteristics of motion, the position and duration are obtained, discussed, and compared. The equivalence of the problems is demonstrated. In section III we conclude.
\par

\section{The two problems}

\subsection{Introduction of the problems}

\begin{flushleft}
The text of the problems is as follows.
\end{flushleft}

\begin{flushleft}
\emph{The Dog-River-Master problem} (DRM)
\end{flushleft}
\par
A dog sees his master on the other side of the river, jumps directly into the water, and swims with a constant speed \(v=|\bold{v}|\) relative to the water, always in the direction of the stationary master. The water has the constant velocity \textbf{w} and the initial distance between them is \textit{L}. How long does it take the dog to reach his master?
\begin{flushleft}
\emph{The Dog-Rabbit-Earth problem} (DRE)
\end{flushleft}
\par
A rabbit runs in a straight line with a constant velocity \textbf{w}. A dog at a constant speed \(v=|\bold{v}|\) starts to pursuit it, always running in the direction of the rabbit. Initially the distance between them is \textit{L}, and \( \textbf{v}\perp\textbf{w}\). Does the dog catch the rabbit?

\par Surveying for the literature we found the DRE problem is intensively studied (see, e.g.,
refs. \cite{20,21,22}, just to mention a few), while for the DRM only some internet references are available. Next, we provide an original solution to the two problems, with a character of generality and easy to follow.

\subsection{The equivalence of the two problems}
\par
To prove the equivalence of the two problems, we start by introducing the cartoon and notations from Figs. 1 and 2. Thus, the trajectory of the dog is sketched in Fig. 1a and 1b relative to Earth (and master, who is at rest relative to Earth, in Figures 1a), and in Fig. 2a and 2b relative to the river or rabbit, respectively. In DRM (see Figs. 1a and 2a) and DRE (see Fig. 1b and 2b) problems, we have: $D\rightarrow $ dog, $R\rightarrow$ river, $M\rightarrow$ master, $E\rightarrow $ Earth, $\textbf{v}\rightarrow $ velocity  of the dog relative to Earth, $\textbf{w}\rightarrow $ velocity of the river or rabbit relative to Earth,$\mathbf{v'}$ $\rightarrow$ velocity of the dog relative to the river or rabbit,  $\textbf{r}\rightarrow$ position vector of the dog relative to Earth, $\mathbf{r'}\rightarrow$ position vector of the dog relative to the river or rabbit. With reference to the kinematics notions, D is the particle, while E, M (identical with E in DRM problem) denotes the \textit{xy}, and R the \textit{x'y'} inertial system of references; at the initial moment E and R are coincident. The similar notations used are meant to suggest the analogies between the two problems.
\par \ \
\par \ \
\par \ \
\par \ \

\begin{figure}[H]
    \includegraphics[width=\textwidth]{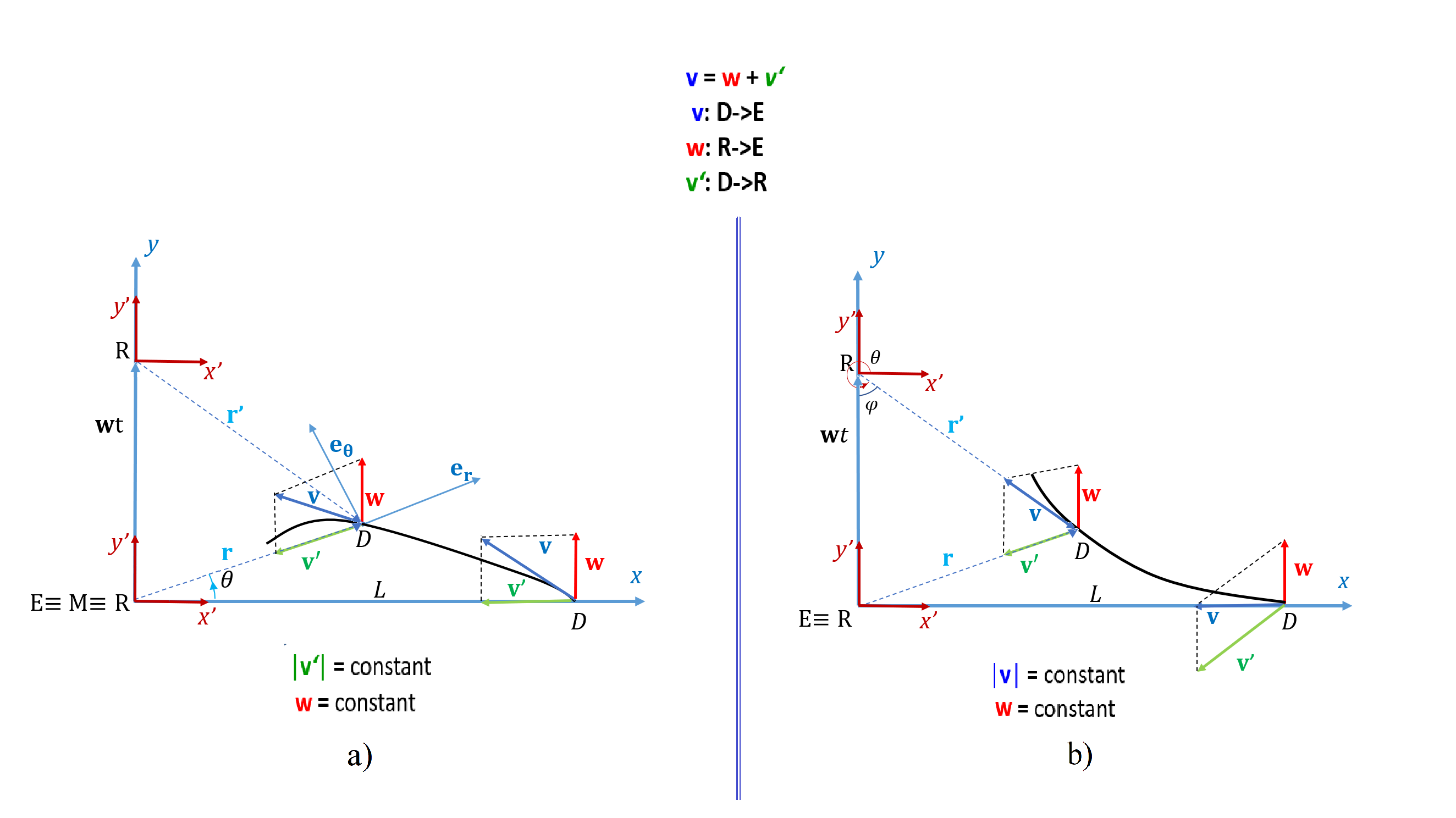}
    \caption{Scheme of the dog trajectory in the \(xy\) frame: a) master's (Earth's) frame (DRM); b) Earth's frame (DRE).}
    \label{fig1}
\end{figure}
\begin{center}
    \includegraphics[width=\textwidth]{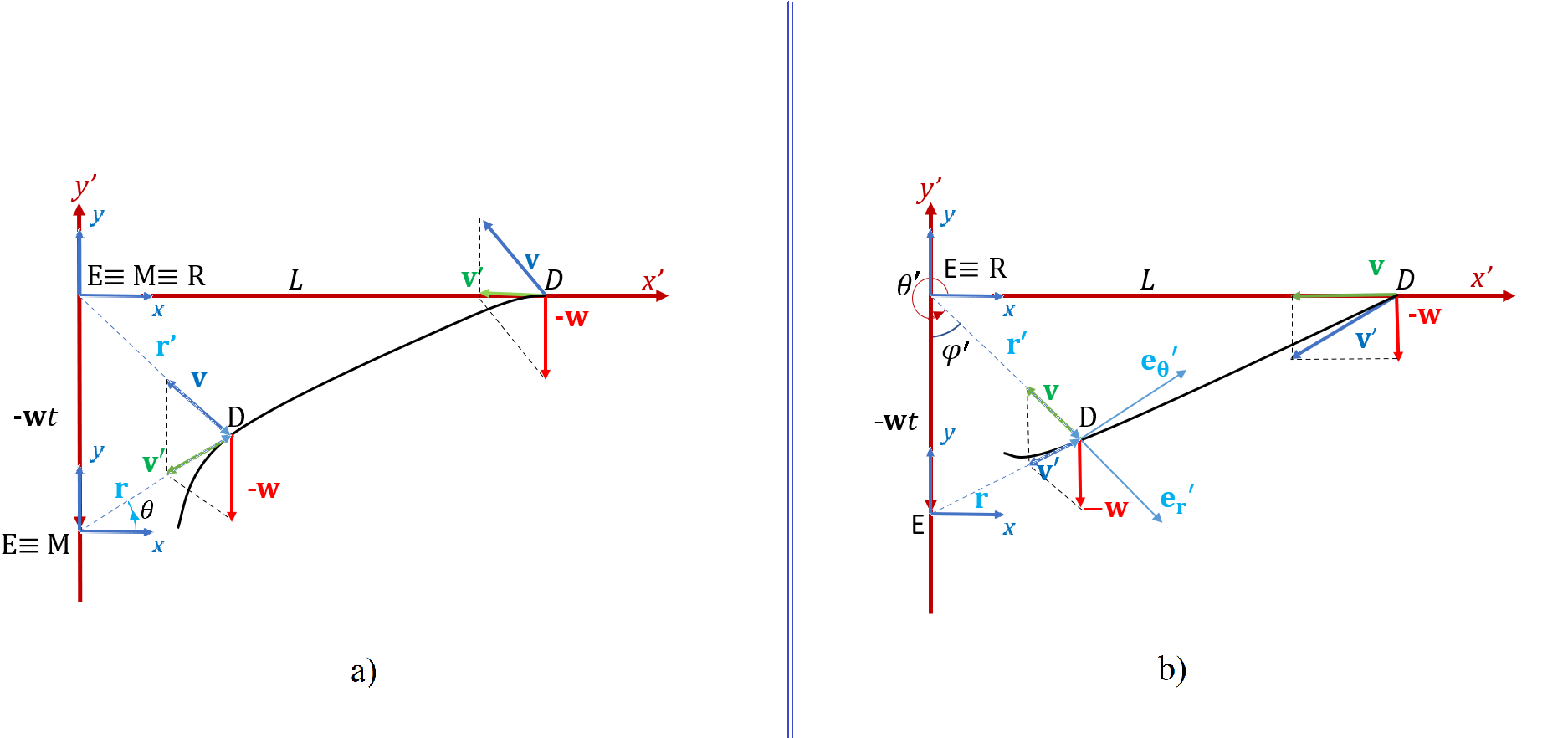}
    \footnotesize Fig. 2-  Scheme of the dog trajectory in the \(x'y'\): a) river's frame (DRM);
b) rabbit's frame (DRE).
\end{center}
\normalsize
\par
\par The vector relations for positions and velocities are as follows:
\begin{equation}
\label{eq1}
    \mathbf{r=w\textit{t}+
    r',}
    \end{equation}
\begin{equation}
    \label{eq2}
    \mathbf{v}=\mathbf{w}+\mathbf{v'}.
\end{equation}

Relation (2) is obtained from eq. (1) by the time derivative operation.
\par
Let's consider first the DRM problem with Figure 1a. By writing the velocity in polar coordinates \(\mathbf{v}=\dot{r}\mathbf{e_r}+r\dot{\theta}\mathbf{e_{\theta}}\) one obtains the system of differential equations:

\begin{equation}
    \label{eq3}
    \left\{\begin{matrix}\dot{r}=-v'+w\sin\theta,\ \\r\dot{\theta}=w\cos\theta.\\\end{matrix}\right.
\end{equation}
\par
\par The DRE problem is solved more efficiently in rabbit's frame (see Fig. 2 b). From eq. (2) by using again the polar coordinates for the velocity \(\mathbf{v'} \), 
that is,\(\mathbf{v'}=\dot{r'}\mathbf{e}_\mathbf{r'}+r'\dot{\theta}\mathbf{e}_\mathbf{\theta'},\) one obtains the system of differential equations to be solved \endflushleft
\begin{equation}
    \label{eq4}
    \left\{\begin{matrix}\dot{r}'=-v+w\cos\varphi'=-v-w \sin\theta',\ \\r'\dot{\theta}'=-w\sin\varphi'=-w \cos \theta',\\\end{matrix}\right.
\end{equation}

\begin{flushleft} where we used \(\varphi'\rightarrow\theta'-3\pi/2\).
Noticing the correspondence between eqs. (3) and (4), namely, \(r\rightarrow r'\) ,\( v\rightarrow v'\), \(\ w\rightarrow-w\) and \(\theta \rightarrow \theta'\), one obtains an unique form of the system of differential equations:\end{flushleft}
\begin{subequations}
    \label{eq5}
    \begin{align}
    \dot{\rho}=-V+W\sin\theta, \label{eq5a} \\
    \rho\dot{\theta}=W\cos\theta, \label{eq5b}
    \end{align}
\end{subequations}

\begin{flushleft}where \(\rho= r\), \(V=v'\), \(W=w\)  for DRM, and \(\rho= r'\),  \(V=v\), \(W=-w\)  for DRE.\end{flushleft}
\par The corresponding initial conditions for the two problems are as follows: 

 \begin{subequations}

  \begin{align}
  r\left(\theta=0\right)=r^\prime\left(\theta'=2\pi\right)=L,\\
 \intertext{and for the case of an affirmative answer to the question of the problems, we have}
   r\left(\theta=\pi/2\right)=r'\left(\theta'=3\pi/2\right)=0.
  \end{align}
\end{subequations}
 
Notice that for the DRM problem \(\theta\) varies with time from 0 to \(\pi/2\)  (increases with time), while according to our choice of axes (see Fig. 1b, 2b) for DRE problem \(\theta'\) varies with time from \(2\pi\) to \(3\pi/2\) (decreases with time).
\subsection{THE RELATIVE MOTION OF THE DOG IN POLAR COORDINATES}

\par For the \textit{dog’s position} in the \textit{xy} (Earth) SR for the DRM problem and in the
\(x' y'\)
(rabbit) SR for DRE problem, we take the ratio of eq. (5a) and eq. (5b), use 
\(\dot{\rho}/\dot{\theta}=d \rho/d \theta\) , do a separation of variables, and obtain 
\begin{equation}
\label{eq7}
\frac{1}{\rho}d\rho=\frac{-V+W\sin\theta}{ W\cos\theta}d\theta.
\end{equation}
\par From the integration of eq. (7) emerges  
\begin{equation}
\label{eq8}
\rho\left(\theta\right)=\frac{C}{\cos{\theta}}\left(\frac{1- \sin\theta}{1+\sin{\theta}}\right)^{ \frac{V}{2W}}=L{\frac{\left(\cos\frac{\theta}{2}-\sin\frac{\theta}{2}\right)^{\frac{V}{W}-1}}{\left(\cos\frac{\theta}{2}+\sin\frac{\theta}{2}\right)^{\frac{V}{W}+1}}},
\end{equation}
\begin{flushleft} where the last equality (which involves \(C=L\)) is obtained from the initial conditions (6a). \end{flushleft}
\par That is:
\begin{center}\(r(\theta)=\frac{L}{\cos{\theta}}\left(\frac{1- \sin\theta}{1+\sin{\theta}}\right)^{ \frac{v'}{2w}}\);
\(r'(\theta')=\frac{L}{\cos{\theta'}}\left(\frac{1- \sin\theta'}{1+\sin{\theta'}}\right)^{ -\frac{v}{2w}}\).
\end{center}

\par For the \textit{duration of the motion}, we firstly add the products of eq. (5a) with \((V + W \sin \theta)\) and of eq. (5b) with \(W \cos \theta\) to obtain\endflushleft
\begin{equation}
\label{eq9}
\dot{\rho}\left(V+W\sin\theta\right)+ \rho \dot{\theta}W\cos\theta=W^2-V^2.
\end{equation}
\par Noticing that the left side equality of eq. (9) can be written as the time derivative of \(\rho (V + W \sin \theta)\), we rewrite eq. (9) as 
\begin{equation}
\label{eq10}
\frac{d}{dt}[\rho (V + W \sin \theta)]=W^2-V^2.
\end{equation}

\par By a separation of variables and an integration from the initial \(\rho\) (equal to \(L\)) and initial \(\theta\) (see eq.(6a)) to an intermediate position characterized by \(\rho(\theta)\)  at time  \(\tau\) , one obtains
\begin{equation}
\label{eq11}
\tau\left(\theta\right)=\frac{\rho\left(V+W\sin\theta\right)-LV}{W^2-V^2},
\end{equation}
\begin{flushleft} with \(\tau(\theta)\equiv t(\theta)\) for DRM and \(\tau(\theta)\equiv t'(\theta')\) for DRE as notations for the time functions.\end{flushleft}
\par The answer to the question of the two problems (if the meeting between the dog and the master or the dog and the rabbit occurs) can be analyzed with eqs. (8) for position and (11) for time. Thus, if:

\begin{flushleft} a) \(v>w \Rightarrow r\left(\theta=\pi/2\right)=r'\left(\theta'=3\pi/2\right)=0\) and the duration is \(t(\theta=\pi/2)=t'(\theta'=3\pi/2)=LV/(V^2-W^2)\);\end{flushleft}

\begin{flushleft} b) \(v=w \Rightarrow r\left(\theta=\pi/2\right)=r'\left(\theta=3\pi/2\right)=L/2\), which is the minimum possible distance in this case. The time until the dog reaches the limit distance \(L/2\) is \(t(\theta=\pi/2)=t'(\theta'=3\pi/2)=\infty\) (see Appendix) and the meeting is not possible;
\end{flushleft}
\begin{flushleft} c) \(v<w \Rightarrow r\left(\theta=\pi/2\right)=r'\left(\theta'=3\pi/2\right)=\infty \) and also from eq. (11) \(t(\theta=\pi/2)=t'(\theta'=3\pi/2)=\infty\) and the meeting again is not possible.

\end{flushleft}

\par Next, several trajectories obtained with Mathematica software are drawn in Fig. 3 by using the Cartesian coordinates \(X=\rho \cos \theta\),  \(Y=\rho \sin \theta\) where
\textit{X, Y} stand for \(x, y\) if \(\rho=r\) or \(x', y'\) if \(\rho=r'\). 
The even character of \(\rho(\theta)\) from eq. (8) relative to simultaneously \(\theta\) and \textit{V} change of sign induces a difference of sign for the \textit{Y}=\textit{Y}(\textit{X}) trajectory function which characterises the two problems.

\par It would be algebraically difficult to write down a Cartesian form of the trajectory of the type \(Y=Y(X)\) by using eq.(8) and the transformation from polar to Cartesian coordinates. This shows the importance of choosing the most appropriate coordinates in order to be able to manage the algebra of the problem.

\begin{center}  \includegraphics[scale=0.25]{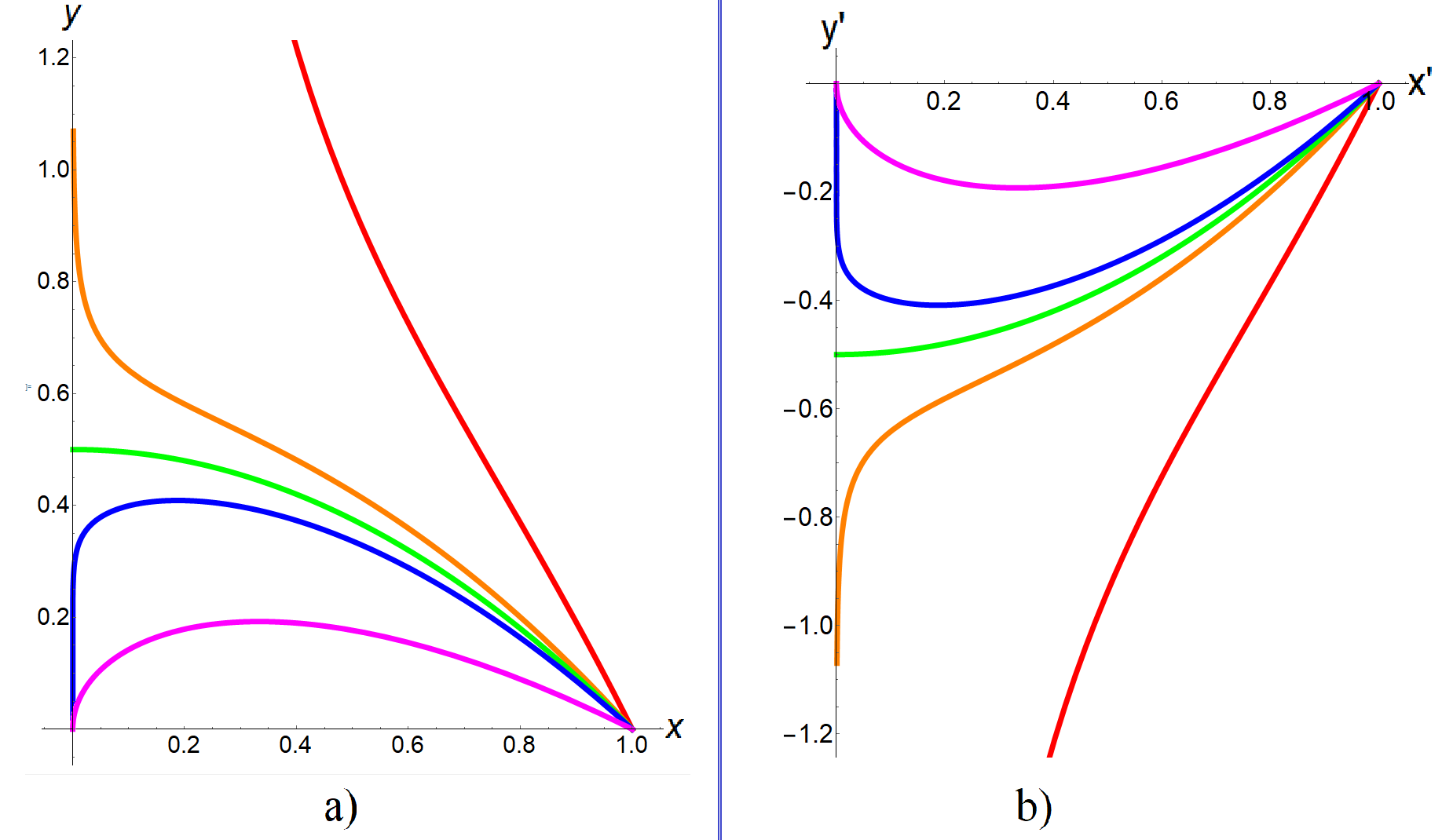}
\flushleft
\centering \footnotesize Fig. 3- the trajectory of the dog in: a) Earth's frame (DRM); b) rabbit's frame (DRE). \end{center}\par
\begin{center} \footnotesize The colors in the graphs correspond to the speed ratios: red for \(v/w=0.5\), orange for \(v/w=0.9\), green for \(v/w=1\), blue for \(v/w=1.1\), magenta for \(v/w=2\); \(L=1\).
\end{center}
\normalsize
\par
\subsection{{THE RELATIVE MOTION OF THE DOG IN CARTESIAN COORDINATES}}
\par For a more comprehensive analysis and a comparison of the results obtained, we 	calculate in this subsection the dog’s trajectory relative to the other SR in each problem. Let’s discuss first the dog’s trajectory relative to Earth (\textit{xy} SR) in the DRE problem. By 	using the position vectors relation (1), from either Fig. 1b or Fig. 2b (notice the angle \(\varphi\) between \(\mathbf{r'}\) and \(\mathbf{-w}\) ), we can write

\begin{equation} 
\label{eq12}
    \left\{\begin{matrix}\ x=r' \sin \varphi'=r' \cos \theta',\\y=wt-r'\cos\varphi'=wt+r'\sin \theta',\\\end{matrix}\right.
\end{equation}

\begin{flushleft} where we used again \(\varphi'=\theta'-3\pi/2\). Introducing notation \end{flushleft}
\begin{center} \(A\equiv L\left(\frac{1+\sin\theta'}{1-\sin{\theta'}}\right)^\frac{v}{2w}\)\end{center}
\begin{flushleft} eq. (12) becomes

\end{flushleft}
\begin{equation}
\label{eq13}
\left\{\begin{matrix}x=A,\\\ y=wt'(\theta')+A\tan\theta'.\\\end{matrix}\right.
\end{equation}

\par From \(x=A\) and with the notation 
\begin{equation}
\label{eq14}
\alpha\equiv\left(\frac{x'}{L}\right)^\frac{v}{2w},
\end{equation}
\begin{flushleft} we obtain
\end{flushleft}

\begin{equation}
    \label{eq}
    \left\{\begin{matrix}\sin{\theta'}=\frac{\alpha-1}{\alpha+1},\ \\\cos{\theta'}=\frac{2\sqrt{\alpha}}{\alpha+1}.\\\end{matrix}\right.
\end{equation}

\par By replacing eq. (8) for the DRE problem and eqs.(14, 15) into eq. (13), one obtains the trajectory of the dog in  \(xy\) SR (the same as in reference \cite{20}): 

\begin{equation}
\label{eq}
y=\frac{1}{2}v\left[\frac{2Lw}{v^2-w^2}+\frac{x\left(\frac{x}{L}\right)^{-\frac{w}{v}}}{w-v}+\frac{x\left(\frac{x}{L}\right)^\frac{w}{v}}{v+w}\right].
\end{equation}

\par In the same manner, that is by using the correspondence mentioned in subsection 2.2, \(x\rightarrow x'\), \(y\rightarrow y'\), \(v\rightarrow v'\),   \(w\rightarrow -w'\),  we can write directly the trajectory of the dog in the river (\(x’y’\) SR). 
Alternatively,  following the same way we obtained eq. (5) from (3) and (4) for velocities, we can write the analogue of eq. (12) for the motion of dog relative to the river (\textit{x'y'} SR) for the DRM problem. Then one can generalize  the expression for the components of positions, and continuing with the algebra as in eqs. (13-15) one obtains a general expression for the Cartesian description of trajectory.
\par Next, several trajectories obtained with Mathematica software are drawn in Figure 4 by using eq. (16) for the DRE problem or its analogue for the DRM problem. Notice that \textit{y}(\textit{x}) from eq. (16) is odd relative to \textit{w} and this is reflected in the changed sign of the trajectories in Figs. 4a and 4b.
\begin{center} \includegraphics[scale=0.165]{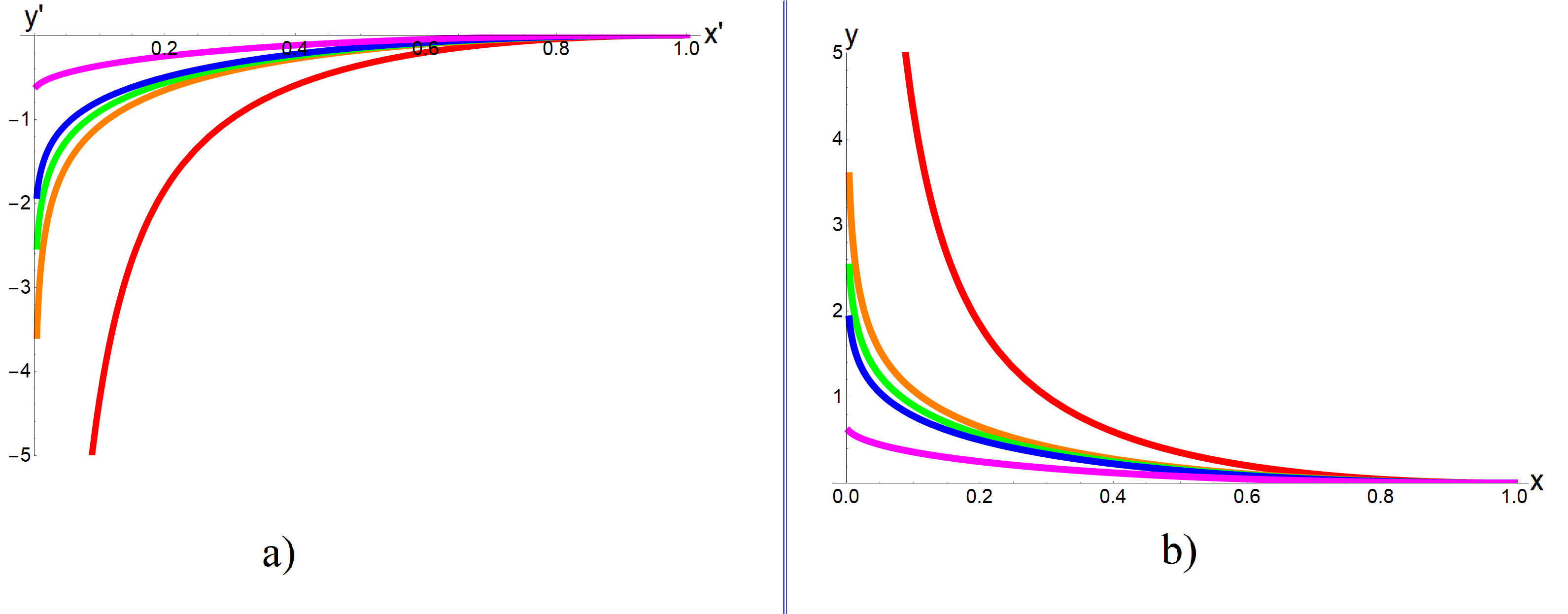}\end{center}
\par
\begin{center} \footnotesize Fig. 4- The trajectory of the dog in: a) river’s frame (DRM); b) Earth’s frame (DRE).\end{center}\par
\begin{center}\footnotesize The colors in the graph correspond to speed ratios: red: \(v/w=0.5\), orange: \(v/w=0.9\), green: \(v/w=1\), blue: \(v/w=1.1\), magenta: \(v/w=2 \).\end{center}

\normalsize

\section{Concluding remarks}
\par We provide a general approach for the DRM and DRE type of kinematics problems. The problems, as they formulate the relative motion of the dog, are proved to be equivalent. The recipe is based on the integration of velocity in polar coordinates in the appropriate SR. Use of the position vectors addition in inertial SRs allows obtaining a Cartesian form for the trajectory. Technically, a key point in developing the algebra is to keep a consistent way in measuring the polar angle  \(\theta\), as we emphasized in Figs. 1 and 2. Implicitly, through eq. (11) and subsequently use of eq. (8)  for 
 \(\rho\left(\theta\right)\), 
our method is also able to provide the time variation of coordinates with time. 
\par The challenging task would be to describe the motion by changing the hypothesis, for example, by a time law given to the motion of the dog. From a pedagogical perspective, we believe that our method would be instructive and of interest to university students studying mechanics.

\section{appendix}
\begin{appendix}
\space
\par For  DRM and  \(v=w\) from eq. (10) we have
\begin{equation}
\label{eq:special} 
\tag{A1} 
 r(\theta)=\frac{L}{1+\sin\theta} 
\end{equation}
and from eq. (4b)
\begin{equation}
\label{eq:special}
\tag{A2}
\int \frac{r(\theta)}{\cos\theta}\,d\theta=u \int dt.
\end{equation}
By integration of eq. (A2), using eq. (A1) and the initial condition \(t(\theta=0)=0\) one obtains 
\begin{center}
\[ t(\theta)=\frac{L}{2u}[\ln{\frac{\cos\theta/2+\sin\theta/2}{\cos\theta/2-\sin\theta/2}}-\frac{1}{(\cos\theta/2+\sin\theta/2)^2}+1] \]
\end{center}
and the meeting time \(t(\theta\rightarrow\pi/2)=\infty\). %\tau\left(\theta->\pi/2\right)=\infty.
\par Similarly, for DRE and v = w from eq. (10) we have
\[
r'(\theta')=\frac{L}{1-\sin\theta'}\tag{A3} \label{eq:special}
\]
and from eq. (4b)
\[ 
-\int \frac{r'(\theta')}{\cos\theta'}\,d\theta'=u \int dt.\label{eq:special}\tag{A4}
\]
By integration of eq. (A4), using eq. (A3) and the initial condition \(t(\theta'=2\pi)=0\) one obtains \[
t(\theta')=\frac{-L}{2u}[\ln{\frac{\cos\theta'/2+\sin\theta'/2}{\cos\theta'/2-\sin\theta'/2}}+\frac{1}{(\cos\theta'/2-\sin\theta'/2)^2}-1]
%\label{eq:special}\tag{A3}
\]
and the meeting time \(t\left(\theta'\rightarrow3\pi/2\right)=\infty\).
\end{appendix}

\end{document}